\documentclass[eqsecnum,aps,tightenlines]{revtex4}
\usepackage{epsfig,graphicx}

\begin{document}

\title{Statistical-Thermal Model Calculations using THERMUS}

\author{S. Wheaton, J. Cleymans}

\address{UCT-CERN Research Centre, Department of Physics, University of Cape Town,\\ Rondebosch 7701, Cape Town, South Africa}

\begin{abstract} 
Selected results obtained using THERMUS, a newly-developed statistical-thermal 
model analysis package, are presented.
\end{abstract}

\maketitle

\section{Introduction}

The statistical-thermal model has proved extremely successful in 
applications to relativistic collisions of both 
heavy ions and elementary particles (cf.~\cite{review} for recent 
surveys and references therein). In light of this success, 
THERMUS~\cite{THERMUS}, a thermal model analysis 
package, 
has been developed for incorporation into the object-oriented 
ROOT framework~\cite{Root}. This 
follows closely on the release of SHARE, a statistical hadronisation 
package, written in \texttt{FORTRAN} and \texttt{Mathematica}~\cite{SHARE}.\\  

At present THERMUS treats the system quantum numbers $B$, $S$ 
and $Q$ within three distinct formalisms: a grand-canonical ensemble, in which 
the quantum numbers are conserved on average; a strangeness-canonical 
ensemble, in which strangeness is exactly conserved, while baryon content and 
charge are treated grand-canonically; and, finally, a fully canonical 
ensemble in which $B$, $S$ and $Q$ are each exactly conserved. A 
detailed theoretical overview of the statistical-thermal model as applicable 
to THERMUS, as well as a description of the structure of THERMUS, can be 
found in~\cite{THERMUS}.\\

\section{Statistical-thermal model predictions using THERMUS}

\begin{figure}
\begin{center}

\includegraphics[width=9cm]{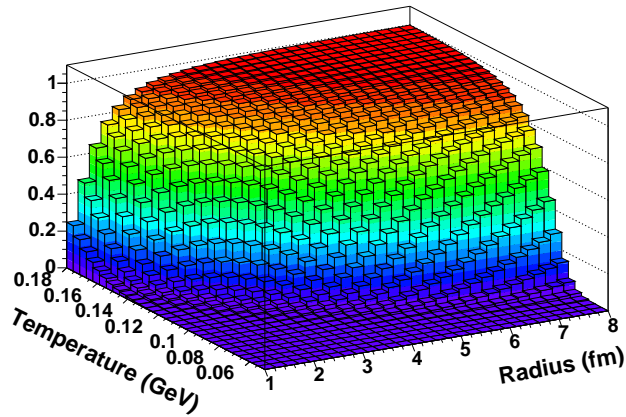}

\end{center}
\caption{The canonical correction factor of a $S=+2$ hadron in the 
strangeness-canonical ensemble as a function of the fireball temperature 
and radius, assuming full strangeness saturation and 
$S_{fireball}=\mu_B=\mu_Q=0$.}\label{CanCorr}
\end{figure}

In a chemical analysis within the grand-canonical ensemble, 
the statistical-thermal model requires 6 parameters as input: the 
chemical freeze-out temperature $T$, baryon-, strangeness- and 
charge chemical potentials $\mu_B$, $\mu_S$ and $\mu_Q$ respectively, 
strangeness saturation factor $\gamma_S$ and fireball volume $V$. 
When restricting one's attention to particle or density ratios, the 
volume dependence largely falls away, unlike in canonical 
treatments where the particle densities within these ensembles differ from those 
in the grand-canonical ensemble by correction factors which are strongly 
volume-dependent~\cite{Keranen}, as seen in Fig.~\ref{CanCorr}.\\


\begin{figure}
\begin{center}

\includegraphics[width=9cm]{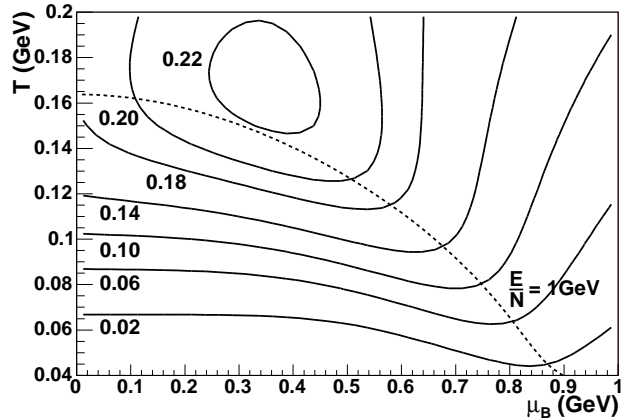}

\end{center}
\caption{Contour plot of the final $K^+/\pi^+$ ratio in the $T-\mu_B$ 
plane with $\mu_S$ constrained to ensure strangeness neutrality, $\mu_Q=0$ and 
$\gamma_S=1$. Also shown as the dashed line is the freeze-out 
condition $E/N=1$ GeV.}\label{KpPipContour}
\end{figure}

Assuming full strangeness saturation (i.e. $\gamma_S=1$) and constraining 
the strangeness density and the ratio $B/2Q$ in the model leaves 
just $T$ and $\mu_B$ as free parameters. This allows all density and 
particle ratios to be plotted in the $T-\mu_B$ plane in which the 
freeze-out condition~\cite{UnifiedFO}, $E/N=1$ GeV, 
defines a curve $T(\mu_B)$ (Fig.~\ref{KpPipContour}).\\ 


\begin{figure}
\begin{center}

\includegraphics[width=9cm]{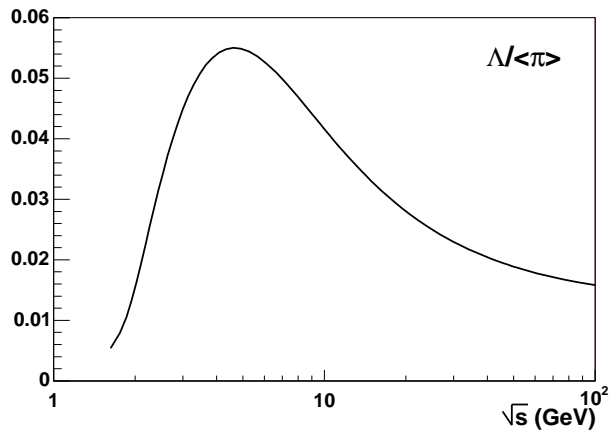}

\end{center}
\caption{The statistical-thermal model prediction for the 
$\Lambda/\left<\pi\right>$ ratio generated by THERMUS.}\label{LaPi}
\end{figure}

\begin{figure}
\begin{center}

\includegraphics[width=9cm]{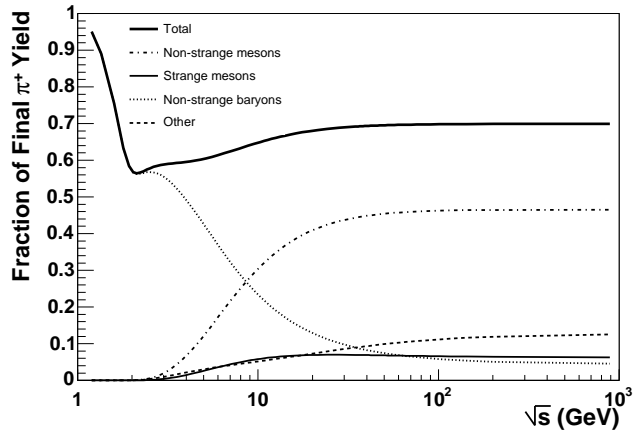}

\caption{The decay contribution to the $\pi^+$ from various 
sources (weak decays excluded).}\label{Pidecay}
\end{center}
\end{figure}

From fits to experimental data it is possible to determine the energy dependence 
of the thermal parameters at chemical freeze-out, as done in~\cite{PBMCleymans}. 
In this way, the statistical-thermal model is able to make predictions as a 
function of collision energy. As an example, in Fig.~\ref{LaPi} the THERMUS 
prediction for the $\Lambda/\left<\pi\right>$ ratio, with $\left<\pi\right>\equiv 
3/2(\pi^++\pi^-)$, is shown. In 
another application, one sees in Fig.~\ref{Pidecay} a breakdown in the feed-down 
contribution to the $\pi^+$. At low energy the 
majority of feeding comes from the baryon sector, while mesons dominate 
at high energy. This highlights the need for a thorough treatment of resonances 
within statistical-thermal models. The THERMUS distribution includes a text  
file listing all hadrons up to mass $2.6$ GeV with $u$, $d$ and $s$ quarks 
listed by the Particle Data Group~\cite{PDG}, as well as text files listing the 
particle decay channels.\\


\section{Thermal fits within THERMUS}

Fits to experimental data are easily accomplished within THERMUS. 
Since feed-down corrections and reconstruction efficiencies 
are often particle- and experiment specific, THERMUS is capable of ascribing to each
particle yield or ratio of interest its own particle set and 
associated decay chain (or separate sets for the numerator and 
denominator in the case of ratios). Use is 
made of the ROOT \texttt{TMinuit} fitting class.\\ 

\begin{figure}
\begin{centering}

\includegraphics[width=9cm]{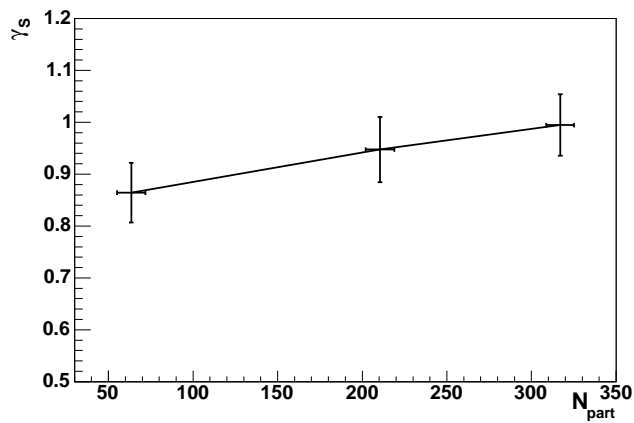}

\end{centering}
\caption{The centrality dependence of the strangeness saturation factor, $\gamma_S$, 
as extracted from Au+Au data at $\sqrt{s_{NN}} = 130$ GeV~\cite{WheatonRHIC}.}\label{Application}
\end{figure}

In a recent application of THERMUS~\cite{WheatonRHIC}, particle ratios measured in Au+Au 
collisions at $\sqrt{s_{NN}} = 130$ GeV by all four 
RHIC experiments, and converted to a common centrality binning in~\cite{Kaneta}, 
were analysed. A grand-canonical treatment was employed with quantum statistics and 
resonance width included for all particles, $\mu_Q=0$, 
%
%
and $T$, $\mu_B$, $\mu_S$ 
and $\gamma_S$ as fit parameters. While the temperature and chemical potentials 
were found to be largely centrality independent, the strangeness saturation factor, 
$\gamma_S$, was observed to increase with participant number towards unity in central 
collisions (Fig.~\ref{Application}). This confirmed previous findings at SPS 
and RHIC energies~\cite{Kaneta,we} that deviations from equilibrium conditions diminish  
with increasing collision centrality.\\ 

\section{Conclusion}

In conclusion, an analysis package, THERMUS, has been developed allowing thermal 
model analyses to be performed within the ROOT framework. Within 
THERMUS statistical-thermal model calculations are possible within 
three distinct formalisms with a complete treatment of hadronic resonances up 
to a mass of $2.6$ GeV. Fits to experimental data are furthermore easily 
achieved.\\

THERMUS is available freely from \texttt{http://hep.phy.uct.ac.za/THERMUS/}, along with 
a detailed user guide and installation instructions.\\

\section{Acknowledgements}

We acknowledge the support of the National Research Foundation (NRF, Pretoria) 
and the URC and PGSO of the University of Cape Town.\\

\end{document}